\documentclass[pre,a4paper,superscriptaddress,showpacs,showkeys,twocolumn]{revtex4-1}
\usepackage{amsmath,amsfonts,amssymb,bm,color,graphicx}
\usepackage{epsfig,grffile,tipa}
\usepackage{soul}
\usepackage{etoolbox}
\usepackage[fixamsmath, disallowspaces]{mathtools}
\usepackage{fouridx}
\usepackage{grfext}
\usepackage[export]{adjustbox}
\DeclarePairedDelimiterX\ket[1]{\lvert}{\rangle}{\mathopen{}#1}
\DeclarePairedDelimiterX\bra[1]{\langle}{\lvert}{\mathopen{}#1}
\DeclarePairedDelimiterX\braket[2]{\langle}{\rangle}%
                               {#1\delimsize\lvert\mathopen{}#2}
\DeclarePairedDelimiterX\ketbra[2]{\lvert}{\lvert}%
                               {\mathopen{}#1\delimsize\rangle\!\delimsize\langle\mathopen{}#2}
\DeclarePairedDelimiterX\Braket[3]{\langle}{\rangle}%
                               {#1\delimsize\lvert\mathopen{}#2\delimsize\lvert\mathopen{}#3}
\DeclarePairedDelimiterX\avg[1]{\langle}{\rangle}%
                               {#1}
\DeclarePairedDelimiterX\norm[1]{\lvert}{\lvert}{\mathopen{}#1}
\DeclarePairedDelimiterX\Norm[1]{\lVert}{\lVert}{\mathopen{}#1}
\DeclarePairedDelimiterX\PP[1]{(}{)}{\mathopen{}#1}
\DeclarePairedDelimiterX\BB[1]{[}{]}{\mathopen{}#1}
\DeclarePairedDelimiterX\CC[1]{\{}{\}}{\mathopen{}#1}

\DeclareGraphicsExtensions{{.png,.pdf,.eps}}

\newcommand{\xalign}{\mathcal{V}}
\newcommand{\xnoise}{\mathcal{S}}

\renewcommand{\vec}[1]{\bm{#1}}
\newcommand{\uvec}[1]{\hat{\bm{#1}}}

\newcommand{\dt}{\Delta t}

\newcommand{\abs}[1]{\left\lvert#1\right\lvert}

\newcommand{\anei}[2]{#2\in\xalign_{#1}}
\newcommand{\xnei}[2]{#2\in\xnoise_{#1}}
\newcommand{\iu}{\imath}
\DeclareMathOperator{\Arg}{arg}

\begin{document}
\title{Noise Induced Phase Separation in Active Systems: Creating
  patterns with noise}
\author{Kosuke Matsui}
\author{John J. Molina} \email{john@cheme.kyoto-u.ac.jp}
\affiliation{Department of Chemical Engineering, Kyoto University, Kyoto 615-8510}
\date{\today}
\begin{abstract}
  We study the flocking and pattern formations of active particles
  with a Vicsek-like model that includes a configuration dependent
  noise term. In particular, we couple the strength of the noise with
  both the local density and orientation of neighboring
  particles. Our results show that such a configuration dependent
  noise can lead to the appearance of large-scale ordered
  and disordered patterns, without the need for any complex alignment interactions. In particular, we obtain an
  ordered band or line state and a disordered active cluster, similar to that seen in the case of
  motility induced phase separation.
\end{abstract}
\maketitle
\section{Introduction}
Understanding the collective dynamics of active systems is currently
one of the most exciting areas of research in soft matter as well as
biological physics. These out-of-equilibrium systems present markedly different
behavior compared to their equilibrium counterparts, such as enhanced
diffusion\cite{Leptos:2009kd}, anomalous
viscosity\cite{Lopez:2015cv}, self-sustained turbulence\cite{Wensink:vs}, giant-number fluctuations\cite{Dey:2012tc}, or
motility-induced phase separation\cite{Cates:2015ft}, to name but a few. Examples of these systems range from the
microscopic scale, including algae, bacteria, and active janus
particles, to the macroscopic scale of fish, birds, buffalo or
humans. The flocking or swarming behavior typically seen in such active systems is usually explained as a
consequence of some type of local velocity aligning interaction,
which is the basis for all Vicsek-like
descriptions\cite{Nagai:2015dj,Mora:2016jh}. Recent works have
challenged this assumption\cite{Romanczuk:2009fl, Ferrante:2013ha,
  Huepe:2014dm, Pearce:2014gs, Barberis:2016gg}, mainly by considering more detailed
cognitive based models, and obtained drastically
different flocking behavior, reminiscent of patterns seen in nature,
such as the marginally opaque flocks of birds\cite{Pearce:2014gs}, the millinglike patterns found in
fish\cite{Gautrais:2012js}, and the file formation seen in
sheep\cite{Toulet:2015kr}. While these approaches have proved
immensely fruitful, they can only be justified for systems where
particle cognition is at play (i.e., not for active janus particles or
synthetic microswimmers).

In this work, guided by the recent investigations into Motility
Induced Phase separation\cite{Cates:2015ft}, as well as the complex
interparticle interactions observed for self-phoretic janus particles\cite{Liebchen:2015baa}
and swimmers in general\cite{Ishikawa:2009et,Llopis:2010in,Zottl:2014wy,Molina:2013hq}, we propose a Vicsek-like model of
flocking that maintains the velocity alignment rule, but that
allows for a configuration dependent noise term. In practice, we
divide the complex particle-particle interactions (which can, for example, be
mediated by the solvent or an additional chemo-attractant)
into an effective alignment together with an asymmetric noise term. With this simple
modification, we are able to observe the emergence of complicated
flocking patterns not seen within the usual Vicsek Model or its
variations. In particular, we find dense polar structures or bands elongated
perpendicular to the direction of propagation, as well as active
disordered clusters, which maintain their size and location in space
over long time scales. Thus, we show that noise can effectively be
used to create large scale patterns in active systems, albeit at the
cost of coupling it to the local density and orientation of
particles. Of interest is the fact that the clustering tendency will
increase both with density and noise, contrary to what is expected for
typical Vicsek-like models.

\section{Model}
\subsection{The Vicsek Model}
The model of active particles proposed by
Vicsek and collaborators\cite{Vicsek:1995eu} describes the off-lattice motion of a set of
$N$ point particles at positions $\vec{x_i}$ ($i=1,\ldots,N$),
which are moving at constant speed $v_0$, in a direction $\theta_i$,
within a 2D periodic domain. For simplicity, it is customary to
represent the velocities as complex numbers, i.e., $\vec{v}_i =
v_0\exp{\left(\iu \theta_i\right)}$, with $\iu$ the imaginary
unit. Driven by the belief that the flocking observed in active systems was due
to a (local) velocity alignment mechanism,  Vicsek et al. proposed the
following set of simplified dynamical rules governing the motion of
the active particles. At each step $t$, every particle will survey its
surroundings, and attempt to align in the average direction of its
neighbors, defined as all the particles within some radius
$R_\xalign$. However, this alignment is not perfect, there is some
noise in the system, and the new orientation $\theta_i(t+\dt)$ will in general
differ from the average alignment the particle has computed. The
updated position $\vec{x}_i(t+\dt)$ is then obtained by having each particle move along
it's new direction during the time interval $\dt$. In this case, it is
assumed that the noise is \textit{intrinsic}\cite{Pimentel:2008en}, it affects all particles
equally and it reflects the fact that all of the particles make some
error during their realignment process (although they are able to
perfectly measure the direction of all their neighbors). Almost ten years after the
original work of Vicsek and his collaborators appeared, Gr\'egoire and
Chat\'e\cite{Gregoire:2004ic}, devised an alternative update scheme, which proposes to
reinterpret the origin of the noise. Instead of assuming an intrinsic
noise, they adopted an \textit{extrinsic} noise source\cite{Pimentel:2008en}, which
represents the inability of the particles to precisely measure the
orientation of their neighbors when deciding upon their new direction
of motion.  Following Pimentel et al.\cite{Pimentel:2008en}, we refer
to the former intrinsic noise model as the \textit{Standard Vicsek Algorithm}
(SVA), and to the latter extrinsic noise model as the
\textit{Gr\'egoire-Chat\'e Algorithm} (GCA). The update rules of both
algorithms can be expressed as\cite{Pimentel:2008en,Chate:2008ca}
\begin{align}
  \theta_i^{\text{SVA}}\PP*{t + \dt} &= \Arg\CC*{\sum_{\anei{i}{j}}
    e^{\iu\theta_j(t)}} + \eta \xi_i(t) \label{e:sva}\\
  \theta_i^{\text{GCA}}\PP*{t + \dt} &= \Arg\CC*{\sum_{\anei{i}{j}} \PP*{e^{\iu\theta_j(t)} + \eta
      e^{\iu\xi_i(t)}}} \label{e:gca} \\
  \vec{x}_i\PP{t + \dt} &= \vec{x}_i\PP{t} + \vec{v}_i{\BB{\theta_i\PP{t+\dt}}}\dt,  
\end{align}
where $\Arg(\vec{z} = z e^{\iu \theta}) = \theta$ is the argument
function, $\xalign_i$ is the alignment region for particle $i$ (of
radius $R_\xalign$), and $\xi_i$ is a delta-correlated white noise random
variable uniformly distributed in $[-\pi, \pi]$, with $\eta\in[0,1]$ the noise
amplitude.

\subsection{The Modified Vicsek Model}
\begin{figure}[ht!]
  \centering
  \includegraphics[width=0.3\textwidth]{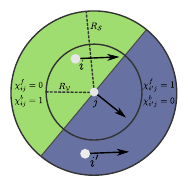}
  \caption{\label{f:xij} (color online) Schematic representation of the alignment and
    noise regions for the \textit{active} noise model. If the distance
    between particles is less than $R_\xalign$ they will contribute to
    each others realignment. In addition, if the distance is less than $R_\xnoise$
    they can contribute to the noise, depending on their relative
    positions and orientations. In the diagram, we consider the
    contributions of a particle $j$ to the updated orientation of
    particles $i$ and $i^\prime$, as given by Eq.~\eqref{e:fba}. In
    the case of ``back'' (``front'') noise, only particles within the
    light green (dark blue) region would feel the noise generated by
    particle~$j$.}
\end{figure}
We propose an alternative variation of the Vicsek model, which again
reinterprets the source of the noise, and find interesting new
dynamical phases. Instead of considering the noise as inherent to the
decision making process of the individual particles, we posit that the
noise can be thought of as arising from the activity of the particles
themselves. As a justification for this interpretation, we can point
to the enhanced diffusion in suspensions of swimming particles\cite{Molina:2014jq}, as
well as the chemorepulsion in active colloidal
dispersions\cite{Liebchen:2015baa}. Although we are working with a
minimal model, which cannot possibly reproduce the detailed dynamics
of such complicated systems, we believe it is possible to incorporate
part of their dynamics within the framework of the Vicsek
model. With a simple rearrangement of Eq.~\eqref{e:gca}, for the
GCA orientation update rule, we obtain
\begin{align}
  \theta_i(t + \dt) &=
  \Arg\CC*{\sum_{\anei{i}{j}}e^{\iu\theta_j(t)} + \eta
    \sum_{\xnei{i}{j}}\chi_{ij}(t )e^{\iu\xi_i(t)}}\label{e:fba}
\end{align}
where we now consider the noise amplitude $\eta \chi_{ij}$ to be
configuration dependent, as well as allow for distinct alignment and
noise regions, $\xalign_i$ and $\xnoise_i$. Here, $\chi_{ij}\in [0,1]$ gives
the relative noise amplitude that particle $j$ generates on particle $i$. It
will depend on the relative positions and orientations of both particles, and will
not be symmetric in general, i.e., $\chi_{ij} \ne \chi_{ji}$. Thus, at each step, a
particle $i$ will realign in the average direction of its neighbors,
located in the alignment region $\xalign_i$ (radius $R_\xalign$);
and this realignment process will exhibit random fluctuations,
generated by some of the particles in the noise region $\xnoise_i$
(radius $R_\xnoise$). We note that in the case where $\chi_{ij} =1$
and $R_\xalign = R_{\xnoise}$ we recover the GCA (Eq.~\eqref{e:gca}).

For simplicity, we will only consider two simple noise functions
$\chi_{ij}^{\text{b}}$ and $\chi_{ij}^{f}$, which we refer to as ``back'' and ``front'' noise
\begin{align}
  \chi_{ij}^{\text{f}}(t) &= H\PP*{\uvec{x}_{ij}\cdot e^{\iu \theta_j(t)}} \label{e:xijf}\\
  \chi_{ij}^{\text{b}}(t) &= H\PP*{\uvec{x}_{ji}\cdot e^{\iu \theta_j(t)}}\label{e:xijb}
\end{align}
where $\vec{x}_{ij} = \vec{x}_i - \vec{x}_j$, carets
($\hat{\cdot}$) denote unit vectors, and $H(x)$ is the Heaviside step
function. Clearly, $\chi_{ij}^{\text{f}}$
($\chi_{ij}^{\text{b}}$) is only different from zero if particle $i$
is located in front (back) of particle $j$. A schematic representation
of the alignment and noise mechanism is given in
Fig.~\eqref{f:xij} for the case where $R_\xnoise > R_\xalign$. Here,
we consider the update of two particles $i$ and $i^\prime$ in the
vicinity of particle $j$. Since $r_{i^\prime j} > R_\xalign$ ($j\notin
\xalign_{i^\prime}$), particle $j$ does not affect the realignment of particle $i^\prime$, it only
contributes to that of particle $i$. However, both particles $i$ and $i^\prime$
can in principle be affected by the noise due to the presence of
particle $j$, since $r_{jk} \le R_\xnoise \quad (k=i,i^\prime)$. This will be
determined by the orientation of particle $j$, the relative positions
of the particles, and the type of noise we are dealing with. In the
case of ``back'' (``front'') noise, only particles at the back (front)
of particle $j$ would experience this noise. This coupling between the
orientation of the particles and the noise will lead to non-trivial
collective behavior which is not seen in either the SVA or GCA
variants of the Vicsek model. 

To characterize the state of our system, we identify the following
five dimensionless parameters using Buckingham's $\Pi$ theorem\cite{Wicks:2007jc,Longair:2003vf}
\begin{align}
  \Pi_\eta  &= \eta \dt &       \Pi_L &= R_\xalign/L \notag\\
  \Pi_v    &= v_0 \dt / R_\xalign & \Pi_R &= R_\xalign/R_\xnoise \label{e:buckpi}\\
  \Pi_\rho &= \pi R_\xalign^2\rho\notag
\end{align}
where $\Pi_\eta$ determines the strength of the noise, $\Pi_v$ gives the
ratio of the distance traveled by a particle in one time step
$v_0\Delta t$ to the alignment radius $R_\xalign$, $\Pi_\rho$ the
average number of particles within the alignment region ($\rho =
N/L^2$ the number density), $\Pi_L$ the ratio of the alignment radius
to the total system size, and $\Pi_R$ the ratio of the alignment
radius to the noise radius. In this work, we are mainly interested in studying how the strength
and type of noise affects the collective properties of the
system; as such, unless otherwise stated, we will focus on the
following region of parameter space: $\Pi_\eta \in (0.0, 1.0)$, $\Pi_R
\in (0.0, 1.0)$, $\Pi_v = 0.01$, $\Pi_\rho = 12.8228$, and $\Pi_L =
0.02857$. As usual when studying swarming of active systems, we
measure the amount of order using the instantaneous orientational order parameter
\begin{align}
  \phi(t) &= \frac{1}{N} \abs{\sum_{i=1}^N e^{\iu \theta_i(t)}}
\end{align}
with $\phi = \avg{\phi(t)}$ the time averaged order parameter.
\section{Results}
\subsection{Phenomenology}
\begin{figure}[ht!]
  \includegraphics[width=0.4\textwidth]{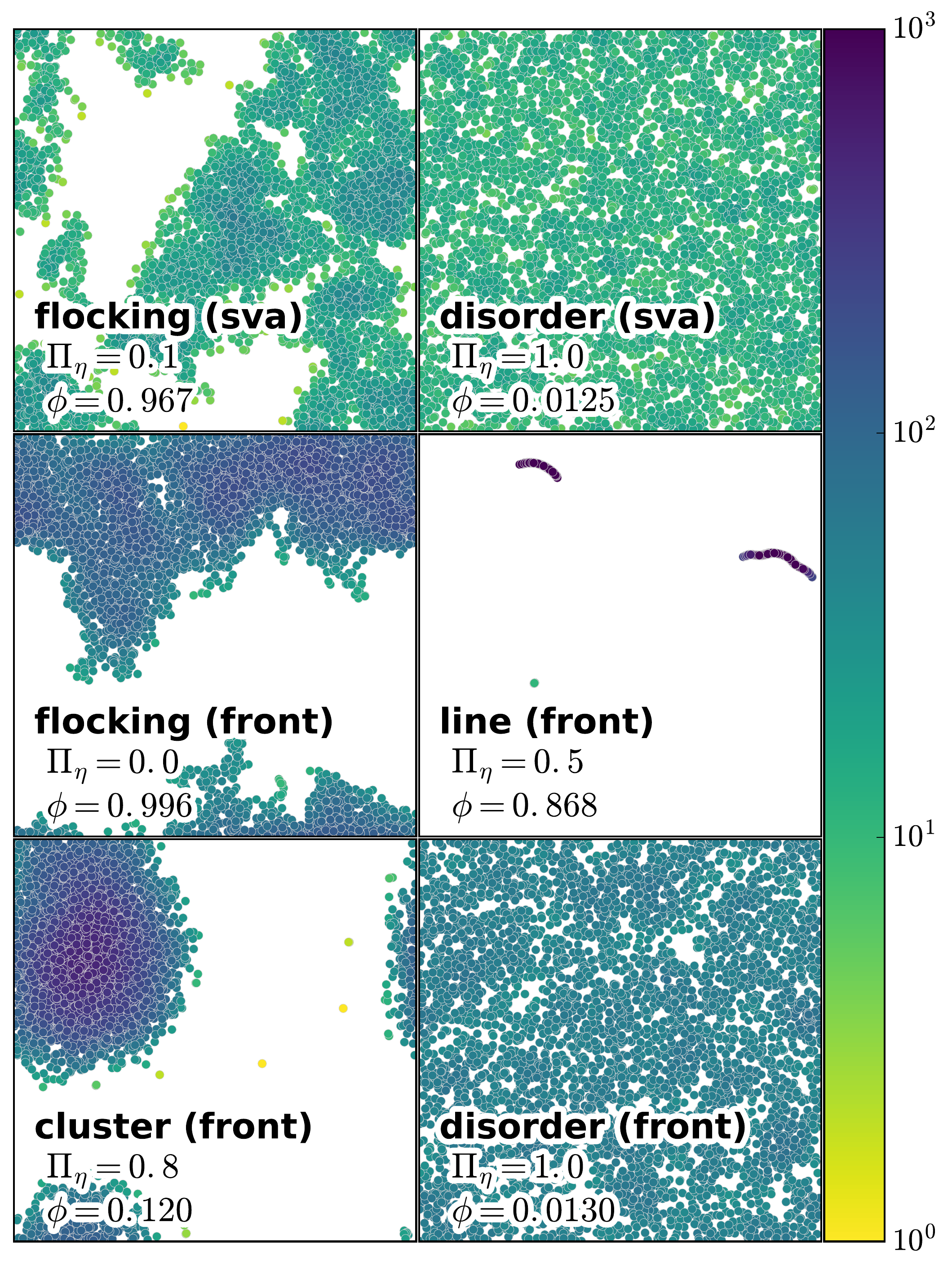}
  \caption{\label{f:snapshots} (color online) Simulation snapshots of the distinct
    dynamical states that can be observed by varying the strength and
    type of the noise. The colormap encodes the local density, i.e, the
    number of neighbor particles within the alignment region $\xalign$
    of each particle. The top panel shows the well known flocking and
    disordered states of the SVA ($\chi_{ij} = 1$ and $\Pi_R =
    1.0$). These can be recovered within the ``front''
    noise model (middle and bottom panels; $\Pi_R = 0.5$), but more interesting are the two
    additional states that appear at intermediate noise strength: a
    moving line state ($\Pi_\eta = 0.5$) and an active disordered cluster state
    ($\Pi_\eta=0.8$).}
\end{figure}
In order to identify the role played by the strength $\Pi_\eta$ and
asymmetry $\Pi_R$ of the noise, we performed simulations over a wide
range of parameters for the two basic models introduced above: the ``front'' and
``back'' noise models. For comparison purposes, we have also carried
out simulations of the SVA under similar parameter ranges. For the
``back'' noise we observe similar behavior to the SVA and will not
discuss this system further. However, in the case of the ``front''
noise particles, we find two novel dynamical states, in addition to
the well-known disorded and ordered/flocking states of the SVA or GCA\cite{Vicsek:1995eu,Chate:2008ca,Pimentel:2008en}: a
disordered active clustering state and an ordered line state. As
expected, for low (high) enough noise intensity $\Pi_\eta$ we recover
an ordered (disordered) state. However, for intermediate noise intensities we see the appearance of dense ordered ($\phi \simeq 1$)
clusters which are elongated perpendicular to the direction of
motion. We refer to these structures as ``lines''. For larger noise
values, such ordered line states are no longer stable. Instead, we see
the appearance of large scale disordered active clusters ($\phi \simeq
0$). Simulation snapshots of these distinct states are given in
Fig.~\ref{f:snapshots}, and the corresponding animated trajectories
are provided as supplemental material\footnote{(Supplemental material) Simulation trajectories
      showing the distinct dynamical states observed for the SVA and
      the front noise model is provided online.}. The appearance of the clusters is of
particular interest, as they are able to maintain their
size and center of mass over relatively large time scales, even though
the constituent particles never stop moving within the cluster and the
cluster itself is constantly exchanging particles with the
environment. Such active clusters have been reported before for active
Brownian particles, both from particle based
simulations\cite{Pohl:2014hn,Tung:2016hq}, as well from a continuum
model with a density dependent noise term\cite{Tailleur:2008kd,Cates:2015ft} (akin to
our configuration dependent noise). In addition, we note that similar states have been observed by
Barberis and Perauni using a minimal flocking model
\cite{Barberis:2016gg} which closely resembles our own variation of
the SVA. However, in Ref.~\cite{Barberis:2016gg}, in contrast
to the present work and most SVA variations, no specific velocity
alignment is included. Instead, a cognitive model is proposed in which
particles reorient using the instantaneous visual information at their
disposal; i.e., the positions of neighboring particles within a
specified visual cone, instead of their velocities. By varying the
size of the vision cone, the authors report the appearance of a line
type state, which they call a worm, an aggregate phase
similar to our active clustering, as well as more complicated
aggregates and nematic bands, which we do not observe. We note
however, that their worms are elongated parallel to the alignment of the
particles, while we see a perpendicular alignment. Nevertheless,
the same type of parallel worms can be obtained using our current model, if the ``front''
noise is replaced with ``left/right'' noise (not shown here).

\subsection{Order Parameter}
\begin{figure}[ht!]
  \includegraphics[width=0.4\textwidth]{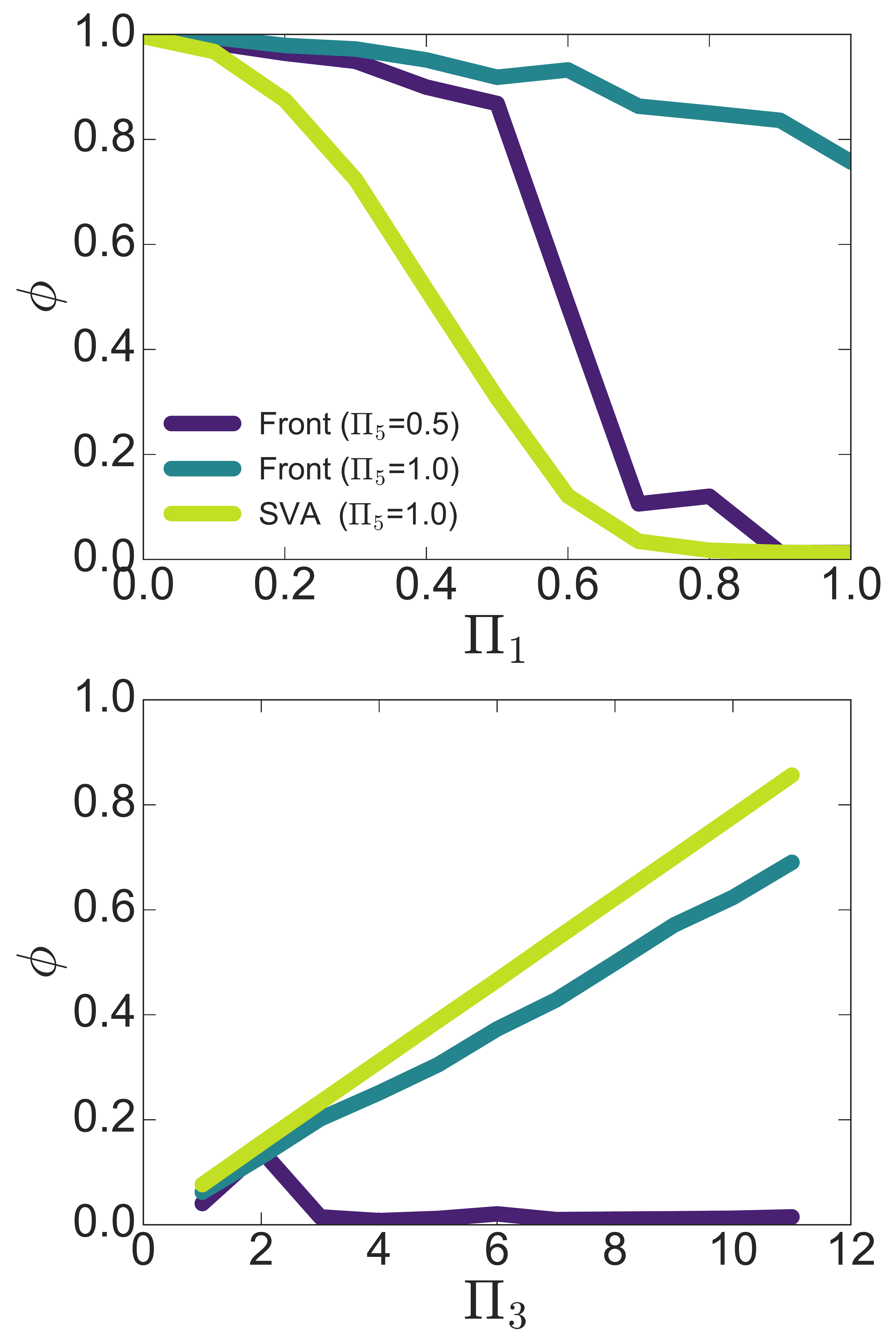}
  \caption{\label{f:orderParam} (color online) Order parameter $\phi$ as a function
    of noise intensity $\Pi_\eta$ (top) and density $\Pi_\rho$
    (bottom, $\Pi_\eta = 0.9$) for two distinct values of the alignment to noise ratio
    $\Pi_R = 0.5, 1.0$. For comparison purposes, we have also included
    the results of the SVA ($\chi_{ij}= \Pi_R = 1$).}
\end{figure}
To study the transition of the system from the low-noise flocking
state to the high-noise disorded state, passing through the line and
clusters, we measured the order parameter $\phi$ as a function of
noise intensity $\Pi_\eta$ for two different alignment to noise size ratios
$\Pi_R = 0.5$~and~$1.0$ (see Figure~\ref{f:orderParam}~a). When the alignment and noise regions
coincide $\Pi_R = 1$ we obtain an ordered phase $\Phi \simeq 1$ for all noise
values. However, the system is not always in a flocking state, as can
be seen by inspecting the trajectories of the system. In fact,
flocking is only stable for very small noise amplitudes $\Pi_\eta
\lesssim 0.1$, for all higher values the stable state is that of the
ordered perpendicular lines. However, if we make the alignment
region smaller than the noise region ($R_\xalign < R_\xnoise$), we
observe a sharp drop in the order parameter at an intermediate noise
$\Pi_\eta \simeq 0.5$. As in the previous case, flocking is only
observed for $\Pi_\eta \simeq 0$; for $0.1\lesssim \Pi_\eta \lesssim 0.5$
the system forms the ordered perpendicular lines, and for higher
values we obtain active disordered clusters. The onset of the
disordered clusters naturally coincides with the noise value at which
the order parameter shows the abrupt drop $\Pi_\eta \simeq 0.5$. We
note that for small to intermediate noise intensities, the dynamics of
the system is insensitive to $\Pi_R$, the differences are only
appreciable for $\Pi_\eta \gtrsim 0.5$. A similar behavior can be seen
for the order parameter as a function of density (see
Figure~\ref{f:orderParam}~b). For low densities, the system is in the
line state for both $\Pi_R = 0.5$~and~$1.0$. As the density is
increased, systems with $\Pi_R = 1.0$ show a slight decrease in order,
while the stable state goes from standard flocking to the
perpendicular line. For $\Pi_R = 0.5$ we observe a sharp drop at
$\Pi_\rho\simeq 3$, at which point the disordered clusters start to develop. The
fact the noise and the density play a similar role can seem
counter-intuitive, particularly since this is not what is seen within
the SVA. However, in our model, noise, alignment, and density are
all coupled, and the appearance of the large scale line and cluster
states is due precisely to the noise, which in turn is caused by the particles
themselves. Therefore, increasing the density has the same net effect as
increasing the strength of the noise.
\subsection{Phase Diagram}
\begin{figure}[ht!]
  \includegraphics[width=0.4\textwidth]{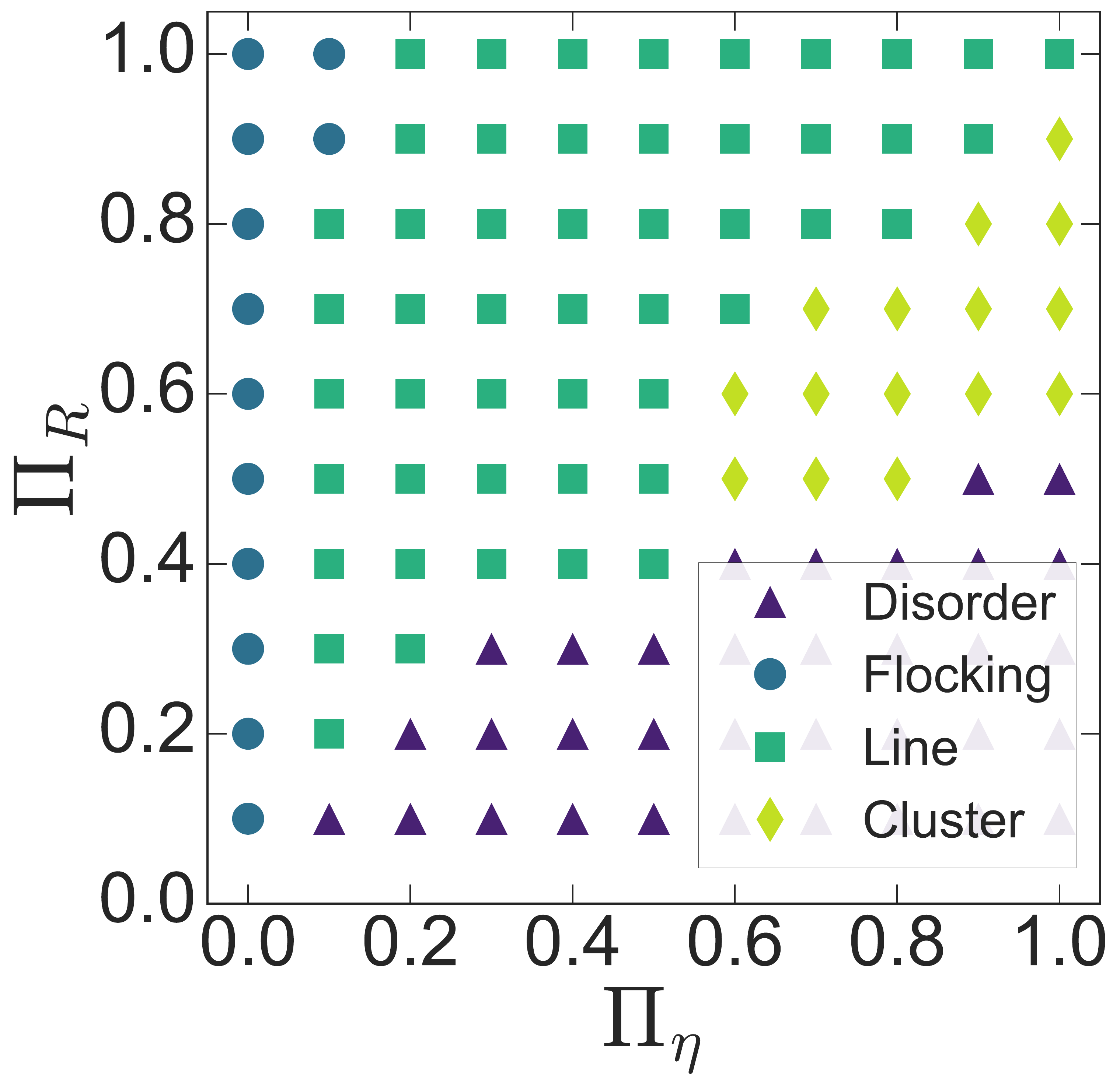}
  \caption{\label{f:phaseDia} (color online) Phase diagram for the forward noise
    systems in the $\Pi_R-\Pi_\eta$ parameter subspace.}
\end{figure}
A detailed summary of the transition between the ordered and
disordered phases is given by the phase diagram shown in
Figure~\ref{f:phaseDia}. We note that even for the highest noise value
$\Pi_\eta = 1$, it is possible to obtain large scale ordered (lines)
as well as disordered (cluster) structures. The homogeneous disordered
phase expected from the SVA only appears for small values of
$\Pi_R$. In addition, we see that the cluster state is only possible
over a narrow range of parameters, for $0.5 \lesssim \Pi_R < 1.0$ and
relatively high noise intensities $\Pi_\eta \gtrsim 0.5$. The line
state seems to be the more stable configuration, as it is 
observed over roughly half the parameter space $\Pi_R > \Pi_\eta >0$.
The fact that the line and cluster states are only observed for
$\Pi_\eta > 0$ is a clear indication that these patterns are induced
by the noise, not by the alignment; although the two are intricately
linked thanks to our interpretation of the former (eq.~\eqref{e:fba}). As
expected, the noise by itself is not enough, we require a certain
degree of alignment, otherwise the system will fall back to
a homogeneous disordered state.

\subsection{Cluster Analysis}
\begin{figure*}[htb!]
  \includegraphics[width=0.45\textwidth,valign=t]{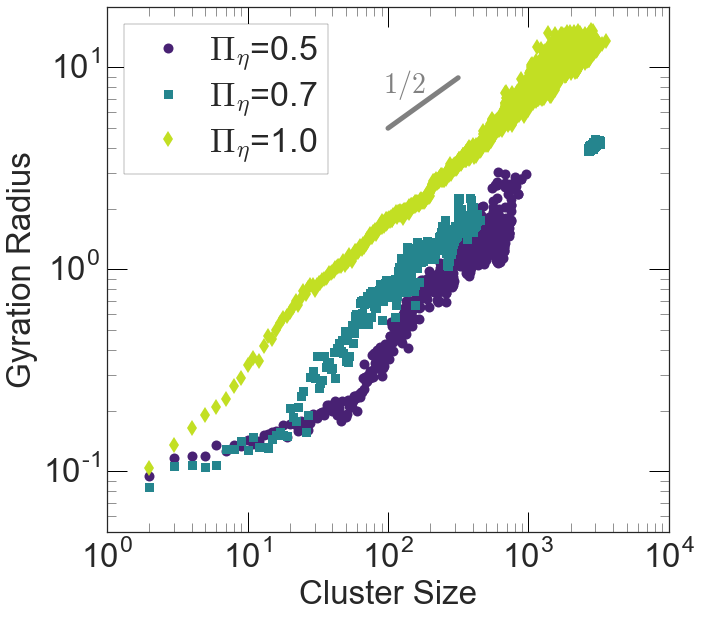}
  \includegraphics[width=0.152\textwidth,valign=t]{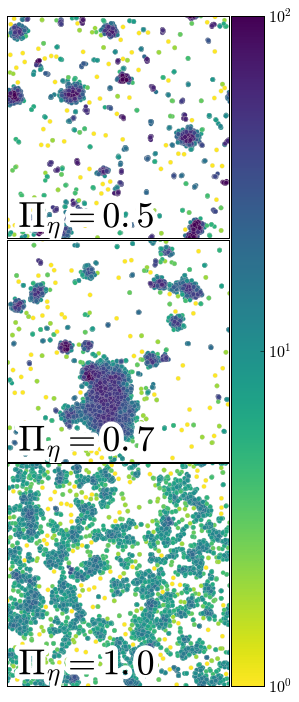}  
  \caption{\label{f:gyration}(color online) (left) Cluster size distribution
    for three different noise values $\Pi_\eta$ and (right) the
    corresponding simulation snapshots, where the color coding represents
    the local particle density (number of particles within the
    alignment region). For these
    simulations, we have used $\Pi_R = 0.3$, $\Pi_v = 0.05$, $\Pi_\rho =
    0.5$, and $\Pi_L = 0.0057$.}
\end{figure*}
Finally, we focus on the appearance of the large scale disordered
clusters. We use a simple distance-based algorithm to
identify the clusters in the system. Thus, if the distance between any
two particles is less than some cutoff distance $r_c$, we consider the
particles to belong to the same cluster. Since we have already
established that it is the noise that is responsible for the
clustering, we choose as cutoff parameter the radius of the noise
region $r_c = R_\xnoise$. Once we have identified the distinct
clusters, we can estimate their size by computing the radius of
gyration 
\begin{align}
  R_g^2 &= \frac{1}{N_c} \sum_{i=1}^{N_c} \left(\vec{x}_i - \avg{\vec{x}}_{\text{c}}\right)^2\label{e:Rg}
\end{align}
with $N_c$ the number of particles in a given cluster and
$\avg{\vec{x}}_c$ its center of mass (average particle
position). In figure~\ref{f:gyration} we show a scatter plot of the
cluster's gyration radius $R_g$, as a function of the cluster size $N_c$
(measured in number of particles), for three different noise values
$\Pi_\eta = 0.5$,~$0.7$,~and~$1.0$ (simulation trajectories are
provided as supplemental material\footnote{(Supplemental material) Simulation trajectories
      showing the active clustering for distinct noise intensities is provided online.}). In the limiting case $N_c\gg 1$,
we recover a power law behavior with exponent $1/2$. This is
equivalent to the gyration radius of an ideal polymer chain and is
further evidence for the entropic origin of the clusters. In addition,
we see that increasing the noise leads to larger clusters. For low
noise values, we obtain relatively small but very dense clusters. As
the noise is increased, the size of the clusters increases, with a
concomitant decrease in the density. For the largest noise intensity,
we obtain a percolating network of broad dilute clusters. Here, to
facilitate the analysis of the cluster formation, we chose a set of
parameters that would give us relatively small clusters (compared to
the single system spanning cluster of Fig.~\ref{f:snapshots}) : $\Pi_R
= 0.3$, $\Pi_v = 0.05$, $\Pi_\rho = 0.5$, and $\Pi_L = 0.0057$. 

\section{Conclusions}
In this work, we proposed a novel variation of the Standard Vicsek
model of active particles, which reinterprets the noise as an
intrinsic quantity, which is coupled to the local particle density and
orientation. Specifically, we consider that in addition to an
alignment in the average direction of its neighbors, each particle will
also ``feel'' a noise which depends on the orientation of its
neighbors. This is in contrast to the usual interpretation of the
noise within such minimal models, which tries to represent an error in
the cognitive process of the particles (i.e., the particles probe
their surroundings and modify their motion accordingly). While the
traditional approach makes sense when one considers the flocking of
animals such as fish, birds, or humans, it is not at all clear how it
can be applied to non-cognitive agents such as active colloidal
particles. When one considers the dynamics of such
self-propelled particles, which can move due to a wide variety of self-phoretic
phenomena, such as diffusiophoresis, electrophoresis, or
thermophoresis, it is obvious that the particles are coupled to their
environment in a highly non-trivial manner. If the particle dynamics
are to be modeled as an effective alignment to neighboring particles
plus a fluctuating noise term, it then makes sense to consider the
noise itself as depending on the local configuration. With this in
mind, we developed a ``forward'' (``backward'') noise Vicsek-like system, in
which the amplitude of the noise felt by any given particle depends on
the number of neighboring particles pointing towards (away) from
it. In spirit, this can be considered as a generalization of the
pusher/puller differentiation of swimming
particles\cite{Molina:2013hq}.

Using the ``forward'' noise model of active particles, we found two
new dynamic states, which are not seen in the standard variations of
the Vicsek model (SVA or GCA):
a highly ordered elongated line/filament, which appears at low to
moderate noise intensities, and an active disordered cluster, which
appears at high noise intensities. The latter is of particular
interest, as it shows how noise can be effectively used to generate
large scale steady patterns. The clusters are a striking example,
as they are composed of moving particles, are constantly exchanging
particles with the environment, and yet are able to maintain
their size and position over large time scales, in the absence of any
external field. In particular, we have shown that it is
the presence of the asymmetric noise that is responsible for the
formation of these active patterns. We have checked the robustness of
the model by using a continuous noise function, instead of the step function
of Eqs.~(\ref{e:xijf}-\ref{e:xijb}), as well as by adding a global
intrinsic noise term (as in \eqref{e:sva}), and a short-range repulsive
interaction.
The same qualitative behavior is obtained; the line and cluster states are still observed,
although the precise boundaries of the phase diagram will of course
vary. To conclude, we have shown how noise can induce large scale patterns in active
systems. We believe this observation can be useful when interpreting
experiments as well as for improving the swarm intelligence of
self-propelled robots, since adding random
fluctuations is easier than computing complex alignment
interactions\cite{Sahin:2005ta}.

\acknowledgments{
  The authors would like to acknowledge Profs. Ryoichi
  Yamamoto and Takashi Taniguchi, Drs. Simon  Schneider and Mitsusuke
  Tarama, and Mr. Norihiro Oyama for valuable
  discussions. This work was supported by the Japan Society for the
  Promotion of Science (JSPS) KAKENHI Grant No. 26247069.
  All figures (except Fig~\ref{f:xij}) were created with the
  \textsf{matplotlib} plotting library\cite{Hunter:ih}.
}

\begin{thebibliography}{33}%
\makeatletter
\providecommand \@ifxundefined [1]{%
 \@ifx{#1\undefined}
}%
\providecommand \@ifnum [1]{%
 \ifnum #1\expandafter \@firstoftwo
 \else \expandafter \@secondoftwo
 \fi
}%
\providecommand \@ifx [1]{%
 \ifx #1\expandafter \@firstoftwo
 \else \expandafter \@secondoftwo
 \fi
}%
\providecommand \natexlab [1]{#1}%
\providecommand \enquote  [1]{``#1''}%
\providecommand \bibnamefont  [1]{#1}%
\providecommand \bibfnamefont [1]{#1}%
\providecommand \citenamefont [1]{#1}%
\providecommand \href@noop [0]{\@secondoftwo}%
\providecommand \href [0]{\begingroup \@sanitize@url \@href}%
\providecommand \@href[1]{\@@startlink{#1}\@@href}%
\providecommand \@@href[1]{\endgroup#1\@@endlink}%
\providecommand \@sanitize@url [0]{\catcode `\\12\catcode `\$12\catcode
  `\&12\catcode `\#12\catcode `\^12\catcode `\_12\catcode `\%12\relax}%
\providecommand \@@startlink[1]{}%
\providecommand \@@endlink[0]{}%
\providecommand \url  [0]{\begingroup\@sanitize@url \@url }%
\providecommand \@url [1]{\endgroup\@href {#1}{\urlprefix }}%
\providecommand \urlprefix  [0]{URL }%
\providecommand \Eprint [0]{\href }%
\providecommand \doibase [0]{http://dx.doi.org/}%
\providecommand \selectlanguage [0]{\@gobble}%
\providecommand \bibinfo  [0]{\@secondoftwo}%
\providecommand \bibfield  [0]{\@secondoftwo}%
\providecommand \translation [1]{[#1]}%
\providecommand \BibitemOpen [0]{}%
\providecommand \bibitemStop [0]{}%
\providecommand \bibitemNoStop [0]{.\EOS\space}%
\providecommand \EOS [0]{\spacefactor3000\relax}%
\providecommand \BibitemShut  [1]{\csname bibitem#1\endcsname}%
\let\auto@bib@innerbib\@empty
\bibitem [{\citenamefont {Leptos}\ \emph {et~al.}(2009)\citenamefont {Leptos},
  \citenamefont {Guasto}, \citenamefont {Gollub}, \citenamefont {Pesci},\ and\
  \citenamefont {Goldstein}}]{Leptos:2009kd}%
  \BibitemOpen
  \bibfield  {author} {\bibinfo {author} {\bibfnamefont {K.~C.}\ \bibnamefont
  {Leptos}}, \bibinfo {author} {\bibfnamefont {J.~S.}\ \bibnamefont {Guasto}},
  \bibinfo {author} {\bibfnamefont {J.}~\bibnamefont {Gollub}}, \bibinfo
  {author} {\bibfnamefont {A.}~\bibnamefont {Pesci}}, \ and\ \bibinfo {author}
  {\bibfnamefont {R.}~\bibnamefont {Goldstein}},\ }\href {\doibase
  10.1103/PhysRevLett.103.198103} {\bibfield  {journal} {\bibinfo  {journal}
  {Physical Review Letters}\ }\textbf {\bibinfo {volume} {103}},\ \bibinfo
  {pages} {198103} (\bibinfo {year} {2009})}\BibitemShut {NoStop}%
\bibitem [{\citenamefont {L{\'o}pez}\ \emph {et~al.}(2015)\citenamefont
  {L{\'o}pez}, \citenamefont {Gachelin}, \citenamefont {Douarche},
  \citenamefont {Auradou},\ and\ \citenamefont {Clement}}]{Lopez:2015cv}%
  \BibitemOpen
  \bibfield  {author} {\bibinfo {author} {\bibfnamefont {H.~M.}\ \bibnamefont
  {L{\'o}pez}}, \bibinfo {author} {\bibfnamefont {J.}~\bibnamefont {Gachelin}},
  \bibinfo {author} {\bibfnamefont {C.}~\bibnamefont {Douarche}}, \bibinfo
  {author} {\bibfnamefont {H.}~\bibnamefont {Auradou}}, \ and\ \bibinfo
  {author} {\bibfnamefont {E.}~\bibnamefont {Clement}},\ }\href {\doibase
  10.1103/PhysRevLett.115.028301} {\bibfield  {journal} {\bibinfo  {journal}
  {Physical Review Letters}\ }\textbf {\bibinfo {volume} {115}},\ \bibinfo
  {pages} {028301} (\bibinfo {year} {2015})}\BibitemShut {NoStop}%
\bibitem [{\citenamefont {Wensink}\ \emph {et~al.}(2012)\citenamefont
  {Wensink}, \citenamefont {Dunkel}, \citenamefont {Heidenreich}, \citenamefont
  {Drescher}, \citenamefont {Goldstein}, \citenamefont {L{\"o}wen},\ and\
  \citenamefont {Yeomans}}]{Wensink:vs}%
  \BibitemOpen
  \bibfield  {author} {\bibinfo {author} {\bibfnamefont {H.~H.}\ \bibnamefont
  {Wensink}}, \bibinfo {author} {\bibfnamefont {J.}~\bibnamefont {Dunkel}},
  \bibinfo {author} {\bibfnamefont {S.}~\bibnamefont {Heidenreich}}, \bibinfo
  {author} {\bibfnamefont {K.}~\bibnamefont {Drescher}}, \bibinfo {author}
  {\bibfnamefont {R.~E.}\ \bibnamefont {Goldstein}}, \bibinfo {author}
  {\bibfnamefont {H.}~\bibnamefont {L{\"o}wen}}, \ and\ \bibinfo {author}
  {\bibfnamefont {J.~M.}\ \bibnamefont {Yeomans}},\ }\href {\doibase
  10.1073/pnas.1202032109} {\bibfield  {journal} {\bibinfo  {journal} {PNAS}\
  }\textbf {\bibinfo {volume} {109}},\ \bibinfo {pages} {14308} (\bibinfo
  {year} {2012})}\BibitemShut {NoStop}%
\bibitem [{\citenamefont {Dey}\ \emph {et~al.}(2012)\citenamefont {Dey},
  \citenamefont {Das},\ and\ \citenamefont {Rajesh}}]{Dey:2012tc}%
  \BibitemOpen
  \bibfield  {author} {\bibinfo {author} {\bibfnamefont {S.}~\bibnamefont
  {Dey}}, \bibinfo {author} {\bibfnamefont {D.}~\bibnamefont {Das}}, \ and\
  \bibinfo {author} {\bibfnamefont {R.}~\bibnamefont {Rajesh}},\ }\href
  {http://prl.aps.org/abstract/PRL/v108/i23/e238001} {\bibfield  {journal}
  {\bibinfo  {journal} {Physical Review Letters}\ }\textbf {\bibinfo {volume}
  {108}},\ \bibinfo {pages} {238001} (\bibinfo {year} {2012})}\BibitemShut
  {NoStop}%
\bibitem [{\citenamefont {Cates}\ and\ \citenamefont
  {Tailleur}(2015)}]{Cates:2015ft}%
  \BibitemOpen
  \bibfield  {author} {\bibinfo {author} {\bibfnamefont {M.~E.}\ \bibnamefont
  {Cates}}\ and\ \bibinfo {author} {\bibfnamefont {J.}~\bibnamefont
  {Tailleur}},\ }\href {\doibase 10.1146/annurev-conmatphys-031214-014710}
  {\bibfield  {journal} {\bibinfo  {journal} {Annual Review of Condensed Matter
  Physics}\ }\textbf {\bibinfo {volume} {6}},\ \bibinfo {pages} {219} (\bibinfo
  {year} {2015})}\BibitemShut {NoStop}%
\bibitem [{\citenamefont {Nagai}\ \emph {et~al.}(2015)\citenamefont {Nagai},
  \citenamefont {Sumino}, \citenamefont {Montagne}, \citenamefont {Aranson},\
  and\ \citenamefont {Chat{\'e}}}]{Nagai:2015dj}%
  \BibitemOpen
  \bibfield  {author} {\bibinfo {author} {\bibfnamefont {K.~H.}\ \bibnamefont
  {Nagai}}, \bibinfo {author} {\bibfnamefont {Y.}~\bibnamefont {Sumino}},
  \bibinfo {author} {\bibfnamefont {R.}~\bibnamefont {Montagne}}, \bibinfo
  {author} {\bibfnamefont {I.~S.}\ \bibnamefont {Aranson}}, \ and\ \bibinfo
  {author} {\bibfnamefont {H.}~\bibnamefont {Chat{\'e}}},\ }\href {\doibase
  10.1103/PhysRevLett.114.168001} {\bibfield  {journal} {\bibinfo  {journal}
  {Physical Review Letters}\ }\textbf {\bibinfo {volume} {114}},\ \bibinfo
  {pages} {168001} (\bibinfo {year} {2015})}\BibitemShut {NoStop}%
\bibitem [{\citenamefont {Mora}\ \emph {et~al.}(2016)\citenamefont {Mora},
  \citenamefont {Walczak}, \citenamefont {Del~Castello}, \citenamefont
  {Ginelli}, \citenamefont {Melillo}, \citenamefont {Parisi}, \citenamefont
  {Viale}, \citenamefont {Cavagna},\ and\ \citenamefont
  {Giardina}}]{Mora:2016jh}%
  \BibitemOpen
  \bibfield  {author} {\bibinfo {author} {\bibfnamefont {T.}~\bibnamefont
  {Mora}}, \bibinfo {author} {\bibfnamefont {A.~M.}\ \bibnamefont {Walczak}},
  \bibinfo {author} {\bibfnamefont {L.}~\bibnamefont {Del~Castello}}, \bibinfo
  {author} {\bibfnamefont {F.}~\bibnamefont {Ginelli}}, \bibinfo {author}
  {\bibfnamefont {S.}~\bibnamefont {Melillo}}, \bibinfo {author} {\bibfnamefont
  {L.}~\bibnamefont {Parisi}}, \bibinfo {author} {\bibfnamefont
  {M.}~\bibnamefont {Viale}}, \bibinfo {author} {\bibfnamefont
  {A.}~\bibnamefont {Cavagna}}, \ and\ \bibinfo {author} {\bibfnamefont
  {I.}~\bibnamefont {Giardina}},\ }\href {http://dx.doi.org/10.1038/nphys3846}
  {\bibfield  {journal} {\bibinfo  {journal} {Nature Physics}\ }\textbf
  {\bibinfo {volume} {12}},\ \bibinfo {pages} {1153} (\bibinfo {year}
  {2016})}\BibitemShut {NoStop}%
\bibitem [{\citenamefont {Romanczuk}\ \emph {et~al.}(2009)\citenamefont
  {Romanczuk}, \citenamefont {Couzin},\ and\ \citenamefont
  {Schimansky-Geier}}]{Romanczuk:2009fl}%
  \BibitemOpen
  \bibfield  {author} {\bibinfo {author} {\bibfnamefont {P.}~\bibnamefont
  {Romanczuk}}, \bibinfo {author} {\bibfnamefont {I.~D.}\ \bibnamefont
  {Couzin}}, \ and\ \bibinfo {author} {\bibfnamefont {L.}~\bibnamefont
  {Schimansky-Geier}},\ }\href {\doibase 10.1103/PhysRevLett.102.010602}
  {\bibfield  {journal} {\bibinfo  {journal} {Physical Review Letters}\
  }\textbf {\bibinfo {volume} {102}},\ \bibinfo {pages} {010602} (\bibinfo
  {year} {2009})}\BibitemShut {NoStop}%
\bibitem [{\citenamefont {Ferrante}\ \emph {et~al.}(2013)\citenamefont
  {Ferrante}, \citenamefont {Turgut}, \citenamefont {Dorigo},\ and\
  \citenamefont {Huepe}}]{Ferrante:2013ha}%
  \BibitemOpen
  \bibfield  {author} {\bibinfo {author} {\bibfnamefont {E.}~\bibnamefont
  {Ferrante}}, \bibinfo {author} {\bibfnamefont {A.~E.}\ \bibnamefont
  {Turgut}}, \bibinfo {author} {\bibfnamefont {M.}~\bibnamefont {Dorigo}}, \
  and\ \bibinfo {author} {\bibfnamefont {C.}~\bibnamefont {Huepe}},\ }\href
  {\doibase 10.1103/PhysRevLett.111.268302} {\bibfield  {journal} {\bibinfo
  {journal} {Physical Review Letters}\ }\textbf {\bibinfo {volume} {111}},\
  \bibinfo {pages} {268302} (\bibinfo {year} {2013})}\BibitemShut {NoStop}%
\bibitem [{\citenamefont {Huepe}\ \emph {et~al.}(2014)\citenamefont {Huepe},
  \citenamefont {Ferrante}, \citenamefont {Wenseleers},\ and\ \citenamefont
  {Turgut}}]{Huepe:2014dm}%
  \BibitemOpen
  \bibfield  {author} {\bibinfo {author} {\bibfnamefont {C.}~\bibnamefont
  {Huepe}}, \bibinfo {author} {\bibfnamefont {E.}~\bibnamefont {Ferrante}},
  \bibinfo {author} {\bibfnamefont {T.}~\bibnamefont {Wenseleers}}, \ and\
  \bibinfo {author} {\bibfnamefont {A.~E.}\ \bibnamefont {Turgut}},\ }\href
  {\doibase 10.1007/s10955-014-1114-8} {\bibfield  {journal} {\bibinfo
  {journal} {Journal of Statistical Physics}\ }\textbf {\bibinfo {volume}
  {158}},\ \bibinfo {pages} {549} (\bibinfo {year} {2014})}\BibitemShut
  {NoStop}%
\bibitem [{\citenamefont {Pearce}\ \emph {et~al.}(2014)\citenamefont {Pearce},
  \citenamefont {Miller}, \citenamefont {Rowlands},\ and\ \citenamefont
  {Turner}}]{Pearce:2014gs}%
  \BibitemOpen
  \bibfield  {author} {\bibinfo {author} {\bibfnamefont {D.~J.~G.}\
  \bibnamefont {Pearce}}, \bibinfo {author} {\bibfnamefont {A.~M.}\
  \bibnamefont {Miller}}, \bibinfo {author} {\bibfnamefont {G.}~\bibnamefont
  {Rowlands}}, \ and\ \bibinfo {author} {\bibfnamefont {M.~S.}\ \bibnamefont
  {Turner}},\ }\href {\doibase 10.1073/pnas.1402202111} {\bibfield  {journal}
  {\bibinfo  {journal} {PNAS}\ }\textbf {\bibinfo {volume} {111}},\ \bibinfo
  {pages} {10422} (\bibinfo {year} {2014})}\BibitemShut {NoStop}%
\bibitem [{\citenamefont {Barberis}\ and\ \citenamefont
  {Peruani}(2016)}]{Barberis:2016gg}%
  \BibitemOpen
  \bibfield  {author} {\bibinfo {author} {\bibfnamefont {L.}~\bibnamefont
  {Barberis}}\ and\ \bibinfo {author} {\bibfnamefont {F.}~\bibnamefont
  {Peruani}},\ }\href {\doibase 10.1103/PhysRevLett.117.248001} {\bibfield
  {journal} {\bibinfo  {journal} {Physical Review Letters}\ }\textbf {\bibinfo
  {volume} {117}},\ \bibinfo {pages} {248001} (\bibinfo {year}
  {2016})}\BibitemShut {NoStop}%
\bibitem [{\citenamefont {Gautrais}\ \emph {et~al.}(2012)\citenamefont
  {Gautrais}, \citenamefont {Ginelli}, \citenamefont {Fournier}, \citenamefont
  {Blanco}, \citenamefont {Soria}, \citenamefont {Chat{\'e}},\ and\
  \citenamefont {Theraulaz}}]{Gautrais:2012js}%
  \BibitemOpen
  \bibfield  {author} {\bibinfo {author} {\bibfnamefont {J.}~\bibnamefont
  {Gautrais}}, \bibinfo {author} {\bibfnamefont {F.}~\bibnamefont {Ginelli}},
  \bibinfo {author} {\bibfnamefont {R.}~\bibnamefont {Fournier}}, \bibinfo
  {author} {\bibfnamefont {S.}~\bibnamefont {Blanco}}, \bibinfo {author}
  {\bibfnamefont {M.}~\bibnamefont {Soria}}, \bibinfo {author} {\bibfnamefont
  {H.}~\bibnamefont {Chat{\'e}}}, \ and\ \bibinfo {author} {\bibfnamefont
  {G.}~\bibnamefont {Theraulaz}},\ }\href {\doibase
  10.1371/journal.pcbi.1002678} {\bibfield  {journal} {\bibinfo  {journal}
  {PLOS Comput Biol}\ }\textbf {\bibinfo {volume} {8}},\ \bibinfo {pages}
  {e1002678} (\bibinfo {year} {2012})}\BibitemShut {NoStop}%
\bibitem [{\citenamefont {Toulet}\ \emph {et~al.}(2015)\citenamefont {Toulet},
  \citenamefont {Gautrais}, \citenamefont {Bon},\ and\ \citenamefont
  {Peruani}}]{Toulet:2015kr}%
  \BibitemOpen
  \bibfield  {author} {\bibinfo {author} {\bibfnamefont {S.}~\bibnamefont
  {Toulet}}, \bibinfo {author} {\bibfnamefont {J.}~\bibnamefont {Gautrais}},
  \bibinfo {author} {\bibfnamefont {R.}~\bibnamefont {Bon}}, \ and\ \bibinfo
  {author} {\bibfnamefont {F.}~\bibnamefont {Peruani}},\ }\href {\doibase
  10.1371/journal.pone.0140188} {\bibfield  {journal} {\bibinfo  {journal}
  {Plos One}\ }\textbf {\bibinfo {volume} {10}},\ \bibinfo {pages} {e0140188}
  (\bibinfo {year} {2015})}\BibitemShut {NoStop}%
\bibitem [{\citenamefont {Liebchen}\ \emph {et~al.}(2015)\citenamefont
  {Liebchen}, \citenamefont {Marenduzzo}, \citenamefont {Pagonabarraga},\ and\
  \citenamefont {Cates}}]{Liebchen:2015baa}%
  \BibitemOpen
  \bibfield  {author} {\bibinfo {author} {\bibfnamefont {B.}~\bibnamefont
  {Liebchen}}, \bibinfo {author} {\bibfnamefont {D.}~\bibnamefont
  {Marenduzzo}}, \bibinfo {author} {\bibfnamefont {I.}~\bibnamefont
  {Pagonabarraga}}, \ and\ \bibinfo {author} {\bibfnamefont {M.~E.}\
  \bibnamefont {Cates}},\ }\href {\doibase 10.1103/PhysRevLett.115.258301}
  {\bibfield  {journal} {\bibinfo  {journal} {Physical Review Letters}\
  }\textbf {\bibinfo {volume} {115}},\ \bibinfo {pages} {258301} (\bibinfo
  {year} {2015})}\BibitemShut {NoStop}%
\bibitem [{\citenamefont {Ishikawa}(2009)}]{Ishikawa:2009et}%
  \BibitemOpen
  \bibfield  {author} {\bibinfo {author} {\bibfnamefont {T.}~\bibnamefont
  {Ishikawa}},\ }\href {\doibase 10.1098/rsif.2009.0223} {\bibfield  {journal}
  {\bibinfo  {journal} {Journal of the Royal Society Interface}\ }\textbf
  {\bibinfo {volume} {6}},\ \bibinfo {pages} {815} (\bibinfo {year}
  {2009})}\BibitemShut {NoStop}%
\bibitem [{\citenamefont {Llopis}\ and\ \citenamefont
  {Pagonabarraga}(2010)}]{Llopis:2010in}%
  \BibitemOpen
  \bibfield  {author} {\bibinfo {author} {\bibfnamefont {I.}~\bibnamefont
  {Llopis}}\ and\ \bibinfo {author} {\bibfnamefont {I.}~\bibnamefont
  {Pagonabarraga}},\ }\href {\doibase 10.1016/j.jnnfm.2010.01.023} {\bibfield
  {journal} {\bibinfo  {journal} {Journal Of Non-Newtonian Fluid Mechanics}\
  }\textbf {\bibinfo {volume} {165}},\ \bibinfo {pages} {946} (\bibinfo {year}
  {2010})}\BibitemShut {NoStop}%
\bibitem [{\citenamefont {Z{\"o}ttl}\ and\ \citenamefont
  {Stark}(2014)}]{Zottl:2014wy}%
  \BibitemOpen
  \bibfield  {author} {\bibinfo {author} {\bibfnamefont {A.}~\bibnamefont
  {Z{\"o}ttl}}\ and\ \bibinfo {author} {\bibfnamefont {H.}~\bibnamefont
  {Stark}},\ }\href
  {http://www.itp.tu-berlin.de/fileadmin/a3233/grk/Veroeffentlichungen/Andreas_2014_PRL.pdf}
  {\bibfield  {journal} {\bibinfo  {journal} {Physical Review Letters}\ }
  (\bibinfo {year} {2014})}\BibitemShut {NoStop}%
\bibitem [{\citenamefont {Molina}\ \emph {et~al.}(2013)\citenamefont {Molina},
  \citenamefont {Nakayama},\ and\ \citenamefont {Yamamoto}}]{Molina:2013hq}%
  \BibitemOpen
  \bibfield  {author} {\bibinfo {author} {\bibfnamefont {J.~J.}\ \bibnamefont
  {Molina}}, \bibinfo {author} {\bibfnamefont {Y.}~\bibnamefont {Nakayama}}, \
  and\ \bibinfo {author} {\bibfnamefont {R.}~\bibnamefont {Yamamoto}},\ }\href
  {\doibase 10.1039/c3sm00140g} {\bibfield  {journal} {\bibinfo  {journal}
  {Soft Matter}\ }\textbf {\bibinfo {volume} {9}},\ \bibinfo {pages} {4923}
  (\bibinfo {year} {2013})}\BibitemShut {NoStop}%
\bibitem [{\citenamefont {Vicsek}\ \emph {et~al.}(1995)\citenamefont {Vicsek},
  \citenamefont {Czir{\'o}k}, \citenamefont {Ben-Jacob}, \citenamefont
  {Cohen},\ and\ \citenamefont {Shochet}}]{Vicsek:1995eu}%
  \BibitemOpen
  \bibfield  {author} {\bibinfo {author} {\bibfnamefont {T.}~\bibnamefont
  {Vicsek}}, \bibinfo {author} {\bibfnamefont {A.}~\bibnamefont {Czir{\'o}k}},
  \bibinfo {author} {\bibfnamefont {E.}~\bibnamefont {Ben-Jacob}}, \bibinfo
  {author} {\bibfnamefont {I.}~\bibnamefont {Cohen}}, \ and\ \bibinfo {author}
  {\bibfnamefont {O.}~\bibnamefont {Shochet}},\ }\href {\doibase
  10.1103/PhysRevLett.75.1226} {\bibfield  {journal} {\bibinfo  {journal}
  {Physical Review Letters}\ }\textbf {\bibinfo {volume} {75}},\ \bibinfo
  {pages} {1226} (\bibinfo {year} {1995})}\BibitemShut {NoStop}%
\bibitem [{\citenamefont {Pimentel}\ \emph {et~al.}(2008)\citenamefont
  {Pimentel}, \citenamefont {Aldana}, \citenamefont {Huepe},\ and\
  \citenamefont {Larralde}}]{Pimentel:2008en}%
  \BibitemOpen
  \bibfield  {author} {\bibinfo {author} {\bibfnamefont {J.~A.}\ \bibnamefont
  {Pimentel}}, \bibinfo {author} {\bibfnamefont {M.}~\bibnamefont {Aldana}},
  \bibinfo {author} {\bibfnamefont {C.}~\bibnamefont {Huepe}}, \ and\ \bibinfo
  {author} {\bibfnamefont {H.}~\bibnamefont {Larralde}},\ }\href {\doibase
  10.1103/PhysRevE.77.061138} {\bibfield  {journal} {\bibinfo  {journal}
  {Physical Review E}\ }\textbf {\bibinfo {volume} {77}},\ \bibinfo {pages}
  {061138} (\bibinfo {year} {2008})}\BibitemShut {NoStop}%
\bibitem [{\citenamefont {Gr{\'e}goire}\ and\ \citenamefont
  {Chat{\'e}}(2004)}]{Gregoire:2004ic}%
  \BibitemOpen
  \bibfield  {author} {\bibinfo {author} {\bibfnamefont {G.}~\bibnamefont
  {Gr{\'e}goire}}\ and\ \bibinfo {author} {\bibfnamefont {H.}~\bibnamefont
  {Chat{\'e}}},\ }\href
  {http://journals.aps.org/prl/abstract/10.1103/PhysRevLett.92.025702}
  {\bibfield  {journal} {\bibinfo  {journal} {Physical Review Letters}\
  }\textbf {\bibinfo {volume} {92}},\ \bibinfo {pages} {025702} (\bibinfo
  {year} {2004})}\BibitemShut {NoStop}%
\bibitem [{\citenamefont {Chat{\'e}}\ \emph {et~al.}(2008)\citenamefont
  {Chat{\'e}}, \citenamefont {Ginelli}, \citenamefont {Gr{\'e}goire},
  \citenamefont {Peruani},\ and\ \citenamefont {Raynaud}}]{Chate:2008ca}%
  \BibitemOpen
  \bibfield  {author} {\bibinfo {author} {\bibfnamefont {H.}~\bibnamefont
  {Chat{\'e}}}, \bibinfo {author} {\bibfnamefont {F.}~\bibnamefont {Ginelli}},
  \bibinfo {author} {\bibfnamefont {G.}~\bibnamefont {Gr{\'e}goire}}, \bibinfo
  {author} {\bibfnamefont {F.}~\bibnamefont {Peruani}}, \ and\ \bibinfo
  {author} {\bibfnamefont {F.}~\bibnamefont {Raynaud}},\ }\href {\doibase
  10.1140/epjb/e2008-00275-9} {\bibfield  {journal} {\bibinfo  {journal} {The
  European Physical Journal B}\ }\textbf {\bibinfo {volume} {64}},\ \bibinfo
  {pages} {451} (\bibinfo {year} {2008})}\BibitemShut {NoStop}%
\bibitem [{\citenamefont {Molina}\ and\ \citenamefont
  {Yamamoto}(2014)}]{Molina:2014jq}%
  \BibitemOpen
  \bibfield  {author} {\bibinfo {author} {\bibfnamefont {J.~J.}\ \bibnamefont
  {Molina}}\ and\ \bibinfo {author} {\bibfnamefont {R.}~\bibnamefont
  {Yamamoto}},\ }\href {\doibase 10.1080/00268976.2014.903004} {\bibfield
  {journal} {\bibinfo  {journal} {Molecular Physics}\ }\textbf {\bibinfo
  {volume} {112}},\ \bibinfo {pages} {1389} (\bibinfo {year}
  {2014})}\BibitemShut {NoStop}%
\bibitem [{\citenamefont {Wicks}\ \emph {et~al.}(2007)\citenamefont {Wicks},
  \citenamefont {Chapman},\ and\ \citenamefont {Dendy}}]{Wicks:2007jc}%
  \BibitemOpen
  \bibfield  {author} {\bibinfo {author} {\bibfnamefont {R.~T.}\ \bibnamefont
  {Wicks}}, \bibinfo {author} {\bibfnamefont {S.~C.}\ \bibnamefont {Chapman}},
  \ and\ \bibinfo {author} {\bibfnamefont {R.~O.}\ \bibnamefont {Dendy}},\
  }\href {http://journals.aps.org/pre/abstract/10.1103/PhysRevE.75.051125}
  {\bibfield  {journal} {\bibinfo  {journal} {Physical Review E}\ }\textbf
  {\bibinfo {volume} {75}},\ \bibinfo {pages} {051125} (\bibinfo {year}
  {2007})}\BibitemShut {NoStop}%
\bibitem [{\citenamefont {Longair}(2003)}]{Longair:2003vf}%
  \BibitemOpen
  \bibfield  {author} {\bibinfo {author} {\bibfnamefont {M.~S.}\ \bibnamefont
  {Longair}},\ }\href
  {https://www.amazon.com/Theoretical-Concepts-Physics-Alternative-Reasoning/dp/052152878X}
  {\emph {\bibinfo {title} {{Theoretical concepts in physics}}}},\ \bibinfo
  {edition} {2nd}\ ed.,\ An alternative view of theoretical reasoning in
  physics\ (\bibinfo  {publisher} {Cambridge University Press},\ \bibinfo
  {address} {Cambridge},\ \bibinfo {year} {2003})\BibitemShut {NoStop}%
\bibitem [{Note1()}]{Note1}%
  \BibitemOpen
  \bibinfo {note} {(Supplemental material) Simulation trajectories showing the
  distinct dynamical states observed for the SVA and the front noise model is
  provided online.}\BibitemShut {Stop}%
\bibitem [{\citenamefont {Pohl}\ and\ \citenamefont
  {Stark}(2014)}]{Pohl:2014hn}%
  \BibitemOpen
  \bibfield  {author} {\bibinfo {author} {\bibfnamefont {O.}~\bibnamefont
  {Pohl}}\ and\ \bibinfo {author} {\bibfnamefont {H.}~\bibnamefont {Stark}},\
  }\href
  {http://gateway.webofknowledge.com/gateway/Gateway.cgi?GWVersion=2&SrcAuth=mekentosj&SrcApp=Papers&DestLinkType=FullRecord&DestApp=WOS&KeyUT=000338020500011}
  {\bibfield  {journal} {\bibinfo  {journal} {Physical Review Letters}\
  }\textbf {\bibinfo {volume} {112}},\ \bibinfo {pages} {238303} (\bibinfo
  {year} {2014})}\BibitemShut {NoStop}%
\bibitem [{\citenamefont {Tung}\ \emph {et~al.}(2016)\citenamefont {Tung},
  \citenamefont {Harder}, \citenamefont {Valeriani},\ and\ \citenamefont
  {Cacciuto}}]{Tung:2016hq}%
  \BibitemOpen
  \bibfield  {author} {\bibinfo {author} {\bibfnamefont {C.}~\bibnamefont
  {Tung}}, \bibinfo {author} {\bibfnamefont {J.}~\bibnamefont {Harder}},
  \bibinfo {author} {\bibfnamefont {C.}~\bibnamefont {Valeriani}}, \ and\
  \bibinfo {author} {\bibfnamefont {A.}~\bibnamefont {Cacciuto}},\ }\href
  {\doibase 10.1039/C5SM02350E} {\bibfield  {journal} {\bibinfo  {journal}
  {Soft Matter}\ }\textbf {\bibinfo {volume} {12}},\ \bibinfo {pages} {555}
  (\bibinfo {year} {2016})}\BibitemShut {NoStop}%
\bibitem [{\citenamefont {Tailleur}\ and\ \citenamefont
  {Cates}(2008)}]{Tailleur:2008kd}%
  \BibitemOpen
  \bibfield  {author} {\bibinfo {author} {\bibfnamefont {J.}~\bibnamefont
  {Tailleur}}\ and\ \bibinfo {author} {\bibfnamefont {M.~E.}\ \bibnamefont
  {Cates}},\ }\href {\doibase 10.1103/PhysRevLett.100.218103} {\bibfield
  {journal} {\bibinfo  {journal} {Physical Review Letters}\ }\textbf {\bibinfo
  {volume} {100}},\ \bibinfo {pages} {218103} (\bibinfo {year}
  {2008})}\BibitemShut {NoStop}%
\bibitem [{Note2()}]{Note2}%
  \BibitemOpen
  \bibinfo {note} {(Supplemental material) Simulation trajectories showing the
  active clustering for distinct noise intensities is provided
  online.}\BibitemShut {Stop}%
\bibitem [{\citenamefont {{\c S}ahin}(2005)}]{Sahin:2005ta}%
  \BibitemOpen
  \bibfield  {author} {\bibinfo {author} {\bibfnamefont {E.}~\bibnamefont {{\c
  S}ahin}},\ }\href {http://link.springer.com/10.1007/978-3-540-30552-1_2}
  {\emph {\bibinfo {title} {{Swarm Robotics: From Sources of Inspiration to
  Domains of Application}}}}\ (\bibinfo  {publisher} {Springer-Verlag},\
  \bibinfo {address} {Berlin},\ \bibinfo {year} {2005})\BibitemShut {NoStop}%
\bibitem [{\citenamefont {Hunter}(2007)}]{Hunter:ih}%
  \BibitemOpen
  \bibfield  {author} {\bibinfo {author} {\bibfnamefont {J.~D.}\ \bibnamefont
  {Hunter}},\ }\href {\doibase 10.1109/MCSE.2007.55} {\bibfield  {journal}
  {\bibinfo  {journal} {Computing in Science {\&} Engineering}\ }\textbf
  {\bibinfo {volume} {9}},\ \bibinfo {pages} {90} (\bibinfo {year}
  {2007})}\BibitemShut {NoStop}%
\end{thebibliography}
%
\end{document}